\documentclass[12pt]{article}

\catcode`\@=11

\def\section{\@startsection{section}{1}{\z@}{3.5ex plus 1ex minus
 .2ex}{2.3ex plus .2ex}{\bf}}

\def\thesubsection{\arabic{section}.\arabic{subsection}}
\renewcommand{\subsection}[1]{\addtocounter{subsection}{1}
\vspace{2.5mm}\par\noindent {\it \thesubsection . #1}\par
 \vspace{0.5mm} }
\catcode`\@=12
\newfont{\mbm}{msbm10}
\def\bb#1{\hbox{\mbm #1}}
\DeclareFontFamily{U}{rsf}{}
\DeclareFontShape{U}{rsf}{m}{n}{
  <5> <6> rsfs5 <7> <8> <9> rsfs7 <10-> rsfs10}{}
\DeclareMathAlphabet\Scr{U}{rsf}{m}{n}

\begin{document}

\begin{titlepage}
\rightline{{CERN-TH/2002-269}}
\rightline{{hep-th/0212066}}
\vskip 2cm
\centerline{{\large\bf Rotating D-branes and O-planes}}
\vskip 1cm
\centerline{Carlo Angelantonj}
\vskip 0.5cm
\centerline{\it Th. Division, CERN  --- CH-1211 Geneva 23}
\vskip  1.0cm
\begin{abstract}
We review orientifold constructions in the presence of 
magnetic backgrounds both in the open and closed sectors. Generically,
the resulting orientifold models have a nice geometric description
in terms of rotated D-branes and/or O-planes. In the case of multiple
magnetic backgrounds, some amount of supersymmetry is recovered if the
magnetic fields are suitably chosen and part of the original D-branes
and/or O-planes are transmuted into new ones.
\end{abstract}
\end{titlepage}

\section{Introduction}

Homogeneous magnetic fields provide an interesting example of a 
non-trivial deformation compatible with two-dimensional conformal 
invariance. The study of their effect in String Theory is relatively
simple for string coordinates are still described in terms of free fields
and the magnetic deformations only affect the world-sheet dynamics by
modifying boundary and periodicity conditions for open and closed
strings.

Despite their simple r\^ole in world-sheet dynamics, magnetic fields lead to
new interesting phenomena in D-branes and orientifold constructions.
Considered at the very beginning as an interesting tool to break 
supersymmetry\cite{witten,bachas}, 
deeper investigations have shown how they can affect the 
geometry of branes and O-planes. As usual, vacuum expectations values 
of gauge fields, after appropriate T-dualities, admit a very elegant and
simple geometrical interpretation: while Wilson lines correspond to brane
displacements, magnetic backgrounds translate into brane/O-plane 
rotations\cite{bdl,ralph,dm,tu,adm}. 
As a result, the corresponding brane/O-plane configurations 
are no longer BPS and typically break supersymmetry. In lower dimensions,
however, some amount of supersymmetry can be recovered if background
fields, {\it i.e.} rotation angles, are chosen appropriately. 

\section{Open-string magnetic field}

Let us start with open-string magnetic fields. As already stated,
the study of their effect is relatively simple, for they 
interact only with the string ends \cite{aboo}. 

For simplicity, let us consider the bosonic string in the presence of a 
uniform magnetic field $H$ in a plane with coordinates $(X^1 , X^2)$.
The variational principle for the world-sheet action
\begin{eqnarray}
{\Scr S} &=& {1\over 2\pi\alpha '}\int {\rm d}\tau \int_0^\pi {\rm d}
\sigma \, \partial_\alpha X^a \partial^\alpha X^a
\\
& & + \left. i q_{\rm L} H \int {\rm d}\tau \, \epsilon_{ab} X^a 
\partial_\tau X^b
\right|_{\sigma=0}
+ \left. i q_{\rm R} H \int {\rm d}\tau \, \epsilon_{ab} X^a \partial_\tau 
X^b \right|_{\sigma=\pi} \,,\nonumber
\end{eqnarray}
yields the wave equation
\begin{equation}
\left( {\partial^2 \over \partial\tau^2} - {\partial^2 \over \partial\sigma^2}
\right) X^a = 0 \,,
\end{equation}
together with the boundary conditions
\begin{eqnarray}
\partial_\sigma X^a - 2\pi \alpha ' q_{\rm L} H \, 
\epsilon^{ab} \partial_\tau X_b &=& 0 \,,
\nonumber \\
\partial_\sigma X^a + 2 \pi \alpha ' q_{\rm R} H \, 
\epsilon^{ab} \partial_\tau X_b &=& 0 \,.
\label{bcond}
\end{eqnarray}

Eq. (\ref{bcond}) admits an alternative geometric formulation in terms of
rotated branes\cite{bdl}. 
Indeed, after performing a T-duality along the $X^2$ direction,
so that $\tau$ and $\sigma$ derivatives are interchanged,
\begin{equation}
\partial_\tau X^2 \rightarrow \partial_\sigma Y^2\,, \qquad
\partial_\sigma X^2 \rightarrow \partial_\tau Y^2\,,
\end{equation}
the boundary conditions become standard Neumann and Dirichlet ones,
say, at $\sigma = 0$
\begin{eqnarray}
\partial_\tau \left( \cos \theta \, X^1 - \sin \theta \, Y^2 \right) &=& 0\,,
\nonumber \\
\partial_\sigma \left( \sin\theta \, X^1 + \cos\theta \, Y^2 \right) &=& 0\,,
\label{rotated}
\end{eqnarray}
for a rotated set of coordinates, where the angle $\theta = \tan^{-1} 
(2 \pi \alpha ' q_{\rm L} H )$
is determined by the magnetic field $H$. 

The interesting feature of intersecting brane constructions that has 
triggered a massive interest is that, in general, chiral fermions
emerge from strings with charged ends while supersymmetry is generically 
broken.

From eq. (\ref{rotated}), however, it is not hard to see that something 
special can happen if one introduces a second magnetic field on a different
two-plane, or, alternatively, if one further rotates the branes along 
another pair of directions. In this case, the choice $\theta_1 = \theta_2 = 
{1 \over 2} \pi$ yields a configuration of orthogonal D7-branes that preserve
half of the supersymmetries. As a result, one can use magnetised, or rotated,
branes to construct, for example, 
new $T^4 /\bb{Z}_2$ type I vacua as in \cite{aads2}.

As a remark, rotated branes are recently playing a pivotal role in brane
world constructions, where the Standard Model degrees fo freedom live at
their intersections \cite{bw}

\section{Compactification on Melvin spaces}

We can now turn to the case of closed-string magnetic backgrounds, 
corresponding to non-vanishing vacuum expectation values for 
the graviphotons\cite{russo1}. 
This background belongs to the class of Melvin metrics
\begin{equation}
{\rm d} s^2 = G_{\mu\nu} (\gamma , R ) {\rm d} X^\mu {\rm d} X^\nu =
{\rm d} \rho ^2 + \rho ^2 \left( {\rm d} \phi + 
{\gamma \over R} {\rm d} y \right)^2 + {\rm d} y^2 + {\rm d} x_i {\rm d} x^i
\,, \label{melvin}
\end{equation}
where $\gamma$ is related to the background magnetic field, 
$(\rho , \phi )$ are polar coordinates on a two-plane, $y$ is the 
compact coordinate on a circle of radius $R$ and $x^i$ are spectator
coordinates.

According to eq. (\ref{melvin}), the string dynamics in the Melvin 
space--time is governed by the world--sheet Action
\begin{equation}
{\Scr S} = - {1\over 4\pi \alpha '} \int {\rm d} \sigma {\rm d} \tau
\, G_{\mu \nu} (\gamma , R) \, \partial_\alpha X^\mu \partial^\alpha X^\nu
\,. \label{strmel}
\end{equation}
This model has a very simple equivalent description after introducing
the complex coordinate
\begin{equation} 
Z = \rho \, {\rm e}^{{\rm i} \varphi} =
\rho \, {\rm e}^{{\rm i} \left( \phi + {\gamma \over R} y\right)} \,,
\end{equation}
that actually corresponds to a free boson
\begin{equation}
\left( {\partial^2 \over \partial \tau^2} - {\partial^2 \over \partial
\sigma^2}\right) Z = 0 
\end{equation}
with the twisted boundary condition
\begin{equation}
Z ( \tau , \sigma + 2 \pi ) = {\rm e}^{2\pi {\rm i} n \gamma} Z (\tau ,
\sigma ) \,,
\label{twist}
\end{equation}
where $n$ is the winding number in the compact $y$ direction.

As a result, we can use all the techniques of orbifold compactifications
to solve exactly the model in eq. (\ref{strmel}).

The background we are considering has the nice property of preserving 
the invariance of the IIB superstring under the world--sheet parity
$\Omega$. In the spirit of \cite{cargese}, we can then proceed to 
construct the associated orientifolds, that, as we shall see, have 
interesting features \cite{adm}.

Aside from the standard $\Omega$ projection, that due to lack of space 
we shall not discuss here (see, however, \cite{adm}), one has 
actually the option
of combining world--sheet parity with other
geometrical symmetries as, for example, parities along the angular 
$\varphi$ and compact $y$ coordinates, $\tilde\Omega = \Omega \Pi_\varphi 
\Pi_y$. The resulting action on the world--sheet coordinates $(y, Z)$ is
\begin{equation} 
\tilde\Omega y (\tau , \sigma ) \tilde\Omega ^{-1}  = - y(\tau , -\sigma) \,,
\qquad \tilde\Omega Z (\tau , \sigma ) \tilde\Omega ^{-1} = Z^\dagger (\tau ,
-\sigma ) \,,
\end{equation}
and, together with the identification (\ref{twist}), implies that 
\begin{equation}
y = - y + 2 \pi s R \,, \qquad \varphi = - \varphi 
+ 2 \pi s \gamma + 2 \pi m \,,
\qquad (s,m \in \bb{Z})
\end{equation}
whose fixed points
\begin{equation}
(y , \varphi )_1 = ( s\pi R , s \pi \gamma ) \,, \qquad 
(y , \varphi )_2 = ( s\pi R , s \pi \gamma + \pi ) \,, \label{fixed}
\end{equation}
identify the positions of orientifold planes. Actually, we are now dealing 
with two infinite sets of rotated O-planes, the angle being proportional 
to the twist, or magnetic background, $\gamma$. Moreover, O-planes in 
different sets undergo a relative $\pi$ rotation. As a result, they carry
opposite R-R charge, a result that we shall soon recover from the 
corresponding Klein-bottle partition function.

In this orientifold construction, although the string partition 
function accounts for the whole spectrum of
physical states, built from the vacuum with string oscillators and/or
Kaluza-Klein modes, a proper derivation of the divergent contributions to
the tree-channel amplitudes is quite involved. In fact, as we have seen, 
a key feature of Melvin compactifications is that the
twist on the complex coordinate $Z$ is coupled to the number of times the
closed string winds the compact coordinate $y$. This means that each
winding state has a different correpsonding ``twisted sector'', and, as such,
in the partition function the characteristics of the theta-functions 
are shifted accordingly. For example, the direct-channel Klein-bottle
amplitude reads
\begin{equation}
{\Scr K} \sim {1\over \tau_{2}^{7/2} \eta^6} \, \sum_n {\Scr K}
(n) \, q^{{1\over 2\alpha'} (nR)^2}\,,
\end{equation}
with
\begin{eqnarray}
{\Scr K} (0) &=& {\textstyle{1 \over 2}} {1\over \tau_2^{1/2} \eta^2}
\sum_{\alpha,\beta} {\textstyle{1\over 2}} 
\eta_{\alpha\beta} {\vartheta^4 \left[
{\alpha \atop \beta} \right] \over \eta^4} \,,
\nonumber \\
{\Scr K} (n) &=& \sum_{\alpha,\beta} {\textstyle{1\over 2}} \eta_{\alpha\beta}
\, {\rm e}^{-{\rm i}\pi \gamma n (2\beta -1)} \, {\vartheta^3 \left[
{\alpha \atop \beta} \right] \over \eta^3} {\vartheta \left[
{\alpha + \gamma n \atop \beta} \right] \over \vartheta \left[
{1/2 + \gamma n \atop 1/2}\right]} \,.
\end{eqnarray}
(Here $\eta_{\alpha\beta} = (-1) ^{2\alpha +2 \beta + 4 \alpha \beta}$
denotes the standard GSO projection.) Therefore, one can not perform 
explicitly the $S$-modular transformation, and the resulting amplitude does
not manifest the usual physical interpretation. 

However, divergent contributions originate only from massless 
states. Hence, to extract their tadpole it suffices to restrict ourselves to
the massless Hamiltonian and to its Kaluza-Klein modification
\begin{equation}
H = {\alpha ' \over 2} \left[ -{1\over \rho} {\partial \over \partial \rho}
\left( \rho {\partial \over \partial \rho} \right) - {1\over \rho^2}
{\partial ^2 \over \partial \phi^2} - {\partial^2 \over \partial y^2}
+ {1\over R^2}\left( k + {\rm i} \gamma {\partial \over \partial \phi}
\right)^2 \right]\,.
\end{equation}
As a result, the leading contributions to the transverse-channel Klein-bottle
amplitude can be extracted from
\begin{equation}
\langle c | {\rm e}^{-\pi\ell H} |c\rangle = \sum_{\Psi} \langle c|\Psi 
\rangle {\rm e}^{-\pi \ell E_\Psi} \langle \Psi | c \rangle\,,
\end{equation}
where $H|\Psi \rangle = E_\Psi |\Psi \rangle$. Taking into account only the 
zero-mode contributions to the boundary state $|c\rangle$, one gets\cite{adm}
\begin{equation}
\langle c | {\rm e}^{-\pi\ell H} |c\rangle_{\rm NS-NS} \sim \sum_{m,k}
\left[1+(-1)^m \right] \left[1+(-1)^k\right] {\rm e}^{-{\pi \alpha ' \ell
\over 2 R^2} (k-\gamma m)^2} \label{NSNS}
\end{equation}
for the NS-NS sector, and
\begin{equation}
\langle c | {\rm e}^{-\pi\ell H} |c\rangle_{\rm R-R} \sim \sum_{m,k}
\left[1-(-1)^m \right] \left[1+(-1)^k\right] {\rm e}^{-{\pi \alpha ' \ell
\over 2 R^2} (k-\gamma m)^2} \label{RR}
\end{equation}
for the R-R sector, where $k$ is the momentum quantum number and $m$
is the quantised angular momentum on the two-plane.
From eq. (\ref{NSNS}) we can then extract the 
non-vanishing dilaton tadpole (for $m=0$ and $k=0$), as well as informations
on the geometry of the O-planes. The factor $1+(-1)^k$ suggests that in
the $y$ direction there are O-planes sitting at $0$ and $\pi R$, while
the projector $1+(-1)^m$ implies that, in $(\rho,\varphi)$ space, there are
pairs of orientifold planes rotated by a $\pi$ angle. Furthermore, from eq. 
(\ref{RR}) one can read that O-planes differing by a $\pi$ rotation have
opposite R-R charge, in agreement with (\ref{fixed}). This phenomenon is
reminiscent of the intersecting brane models discussed in the previous section.
Although the overall R-R tadpole vanishes, cancellation of the NS-NS one calls
for the introduction of D-branes, as in standard orientifold constructions
\cite{adm}.

Also in this case, one can generalise the construction considering multiple
two-planes subject to independent twists. We have now the option to couple
each twist to the same $S^1$ or to different ones. While the latter choice
corresponds to a trivial generalisation of the single twist previously 
reviewed, the former turns out to be quite interesting and leads to amusing
phenomena. 

To be more specific let us consider the case of two 
two-planes with twists $\gamma_1$ and $\gamma_2$. A simple analysis of Killing
spinors on this Melvin background\cite{russo2} 
shows that whenever the $\gamma_i$ are even 
integers all supersymmetry charges are preserved. However, one has the 
additional possibility of preserving only half of the original 
supersymmetries if
\begin{equation}
\gamma_1 \pm \gamma_2 \in 2 \bb{Z} \,.
\end{equation}
This is very reminiscent of the condition one gets for orbifold 
compactifications. Modding out by $\Omega \Pi_y \Pi_{\varphi_1} \Pi
_{\varphi_2}$ then introduces O-planes at the fixed points of the 
involution. Particularly interesting is the case $\gamma_i = 
{1\over 2}$. The resulting model is the traditional $(\bb{C}^2 \times S^1
) /\bb{Z}_2$ orbifold, where the $\bb{Z}_2$ acts as a reflection on the
$\bb{C}^2$ coordinates accompanied by a momentum or a winding shift along
the compact circle, and mutually orthogonal O6 planes are 
generated\cite{adm}, as
pertains to a $(\bb{C}^2 \times S^1) /\bb{Z}_2$ orientifold.

\vskip 24pt

\noindent
{\bf Acknowledgments.} It is a pleasure to thank the organisers 
of the first SP2002 conference, and of the
$35^{\rm th}$ International Symposium Ahrenshoop on the  
Theory of Elementary Particles for their kind invitation.

\end{document}